\pdfoutput=1

\documentclass{vgtc}                          
\UseRawInputEncoding

\graphicspath{{figures/}{pictures/}{images/}{./}} 

\usepackage{times}

\usepackage{tabu}                      
\usepackage{booktabs}                  
\usepackage{lipsum}                    
\usepackage{mwe}                       

\usepackage{mathptmx}                  
\usepackage{authblk}
\usepackage{helvet}  

\usepackage{multirow}
\usepackage{makecell}

\onlineid{0}

\vgtccategory{Research}

\vgtcinsertpkg

\title{``I Feel Myself So Small!'': Designing and Evaluating VR Awe Experiences Based on Theories Related to Sublime}

\author{
    {\fontsize{10pt}{12pt}\selectfont \textsf{Zhiting He}$^{*}$\thanks{*\ e-mail: \textnormal{\{zhiting, mfan, Gxinyi, t\_90800\}@cuc.edu.cn}}} \hspace{1em}
    {\fontsize{10pt}{12pt}\selectfont \textsf{Min Fan}$^{*\dag}$\thanks{\dag\ corresponding author}} \hspace{1em}
    {\fontsize{10pt}{12pt}\selectfont \textsf{Xinyi Guo}$^{*}$} \hspace{1em}
    {\fontsize{10pt}{12pt}\selectfont \textsf{Yifan Zhao}$^{\ddag}$\thanks{\ddag\ e-mail: yz4577@columbia.edu}} \hspace{1em}
    {\fontsize{10pt}{12pt}\selectfont \textsf{Yuqiu Wang}$^{*}$}
}

\affil{\vspace{-0.5em}\small\fontsize{8pt}{10pt}\selectfont\textsf{
    $^{*}$Communication University of China \hspace{0.5cm} $^{\ddag}$Columbia University in the City of New York
}}

\makeatletter
\renewcommand\@fnsymbol[1]{}
\makeatother

\abstract{
Research suggests the potential of employing VR to elicit awe experiences, thereby promoting well-being. Building upon theories related to the sublime and embodiment, we designed three VR scenes to evaluate the effectiveness of sublime and embodied design elements in invoking awe experiences. We conducted a within-subject study involving 28 young adults who experienced the three VR designs. Results demonstrated that the VR design with sublime elements significantly elicited more intense awe experiences compared to the one without, while adding embodied elements did not enhance the intensity of awe. Qualitative interviews revealed critical design elements (e.g., the obscure event should be reasonable) and their underlying mechanisms (e.g., leading to feelings of enlightenment) in invoking awe experiences. We further discuss considerations and implications for the design of effective awe-inspiring VR applications. 
} 


\begin{document}

\maketitle

\section{Introduction}
Recent research suggests the potential of utilizing interactive technology to support self-transcendent positive emotional experiences, thereby promoting physical and psychological well-being~\cite{calvo_positive_2014,moreira-almeida_wpa_2016,mossbridge_designing_2016}. Awe is a positive emotion that emerges when individuals encounter something greater than themselves~\cite{kitson_designing_2020}, and it has been shown to foster prosocial behaviors~\cite{monroy_awe_2022,pizarro_self-transcendent_2021}, pro-environmental behaviors~\cite{hicks_learning_2020}, and reduced anxiety~\cite{monroy_influences_2023,valdesolo_awe_2014}. Awe-inducing stimuli include magnificent natural scenes, displays of human courage, ceremonial activities, and more~\cite{monroy_awe_2022}. Individuals can access these phenomena either through direct experiences or indirectly by observing them in books, films, and other mediums. However, compared to direct experiences, the indirect method is often regarded as a passive elicitation, potentially diminishing the intensity of emotional responses~\cite{somarathna_virtual_2022}. To enable individuals to have more direct experiences of awe when they lack the time and resources for real-life encounters, virtual reality (VR) offers a promising active emotional induction method. VR has been demonstrated to be as effective as reality in eliciting intense experiences and emotions~\cite{somarathna_virtual_2022,miller_awedyssey_2023,riva_neuroscience_2019}.

However, there has been a noticeable absence of a systematic design rationale for evoking awe through VR. Currently, some studies rely on observational research of awe-related natural scenes and human experiences, providing specific design features that may not be easily transferred and applied to other design cases. These studies often lack integrated theories to provide a solid rationale for their design suggestions~\cite{kim_designing_2022}. While Keltner and Haidt suggested two central appraisals for awe, namely \textbf{\textit{perceived vastness}} and \textbf{\textit{a need for accommodation}}~\cite{keltner_approaching_2003}, these features are primarily related to emotional experience and cannot be directly translated into design principles~\cite{miller_awedyssey_2023,keltner_approaching_2003}. To bridge this gap, we turn to systematic theories of the sublime, which are suggested to be closely related to the awe experience according to Clewis et al.'s study~\cite{clewis_intersections_2022}. Therefore, we hypothesized that designing sublime experiences could evoke a sense of awe. Drawing design-related elements from theories of the sublime, we aimed to create VR sublime experiences and subsequently conducted an empirical study to examine its effectiveness in inducing awe.

Additionally, research has demonstrated that the sense of embodiment can impact the intensity of emotions in VR experiences, such as anxiety~\cite{chen_how_2017}, and it can alter participants’ perceived size and distance of surrounding objects in VR~\cite{ries_effect_2008,ogawa_virtual_2019,ebrahimi_investigating_2018}. Therefore, we hypothesized that embodiment might also influence the intensity of awe, an aspect that has not been fully explored in current VR awe design studies. Thus, our research questions are: \textbf{RQ1:} Can design elements drawn from theories related to the sublime evoke feelings of awe in participants, and how? \textbf{RQ2:} Can embodied experience enhance participants’ intensity of awe, and how?

We created three immersive VR scenes to elicit the awe experience: one without either sublime or embodied design elements, one with sublime design elements only, and one with both sublime and embodied design elements. Next, we conducted a within-subject study with 28 young adults who experienced the three VR designs. The results demonstrated that the VR design with sublime elements significantly elicited more intense awe experiences compared to the scene without, while the addition of embodied elements alongside sublime one did not further enhance the intensity of the awe experience compared to scenes with sublime elements alone. Qualitative interviews provided additional insights into the effects and mechanisms of sublime and embodied design elements. We further discuss considerations and implications for the design of effective awe-inspiring VR applications.

Our research offers three main contributions. (1) We discussed the systematic design rationale for awe-inspiring VR design based on theories related to the sublime. (2) We presented a proof-of-concept VR design based on the rationale that uses sublime design elements and embodied design elements to induce awe in young adults. (3) We conducted empirical studies and explicated the effect and mechanism of sublime and embodied design elements in inducing awe. The knowledge may benefit designers and researchers who are interested in creating effective awe-inspiring VR experiences in the future.

\section{Related Work and Theoretical Foundations}

\subsection{Theories of Sublime}

Recent research has found a strong link between awe and the sublime and generally suggested their significant overlap~\cite{miller_awedyssey_2023, arcangeli_awe_2020, graves_friluftsliv_2020}. In discussing the relationship between awe and the sublime, Gordon et al.~\cite{gordon_dark_2017} proposed that the sublime is a concept related to threat-based awe, with both threat and fear being crucial to these experiences. Arcangeli et al.~\cite{arcangeli_awe_2020} pointed out that both awe and the sublime are mixed valence experiences, involving both positive and negative evaluations. However, they distinguished between the two by stating that the sublime is inherently an aesthetic experience, whereas awe does not necessarily have to be one. Therefore, the positive evaluation present in awe, such as admiration, is not a key concept in the experience of the sublime. Clewis et al.~\cite{clewis_intersections_2022} also considered the sublime as a form of aesthetic awe and highlighted several similarities between awe and the sublime. These include the characterization of both as mixed valence experiences that are overall pleasant, the involvement of a perception of vastness, and a need for accommodation. Moreover, both experiences require maintaining a certain distance from the object or event, with the experiencer adopting an observational stance rather than being actively involved. 

The overlap between sublime experiences and awe experiences might inspire the design of awe experiences. Therefore, we systematically reviewed the philosophical literature on the sublime and analyzed the characteristics of sublime experiences that are similar to awe experiences. Regarding the source of the sublime, Longinus~\cite{longinus_sublime_1995} believed that it should be awe-inspiring rather than merely pretty, as “pretty" dismisses the element of danger. Burke~\cite{burke_philosophical_2019} proposed that \textbf{\textit{terror}}, or anything inherently terrifying, is a source of the sublime. This includes the terror of danger, power, infinity, or great extremes of dimension. Similarly, Schopenhauer~\cite{schopenhauer_world_1969} pointed out that the contemplated object should have an unfavorable and hostile relation to the will while Kant~\cite{kant_critique_2009} added that what elicits the feeling of the sublime may,  in its form, appear contrary to our power of judgment. Burke further emphasized that terror should be kept at a certain \textbf{\textit{distance or modified}} so that viewers can derive pleasure from the horrifying situation. Kant similarly noted that for someone to continue feeling sublime, they must experience the cessation of a troublesome situation and avoid that danger again. Schopenhauer~\cite{schopenhauer_world_1969} argued that actual personal affliction and danger disrupt peace and contemplation, diminishing the sublime as anxiety takes over. In addition to terror, Burke~\cite{burke_philosophical_2019} suggested that \textbf{\textit{obscurity}} is a significant feature of sublime objects, preventing viewers from fully comprehending the extent of terror. Kant~\cite{kant_critique_2009} also highlighted that the feeling of the sublime is not expressed through specific forms but rather through the intentional use of these forms by the imagination. 

Building on previous theorists' perspectives, we adopt the idea that the sublime experience involves elements of \textbf{\textit{terror}}, which can stem from either fearful objects or objects of great magnitude. Due to \textbf{\textit{obscurity}}, viewers should find fearful objects unfamiliar, and objects of magnitude incomprehensible. As a result, these elements can create an unnatural tension, causing discomfort and triggering self-preservation, leading viewers to try to apprehend the situation. However, when presented at a certain \textbf{\textit{distance or with modifications}}, such as in a safe environment or a comprehensible context, terror can evoke delight. Therefore, when people have an idea of pain, without being actually in such circumstances, they feel sublime.

\subsection{Theories of Embodiment}

The sense of agency and ownership has been found to influence higher-level emotional experiences through bodily self-perception~\cite{chen_how_2017}. Virtual avatars have the potential to serve as a reference point, altering participants’ perception of surrounding objects in VR. For instance, Ries et al.~\cite{ries_effect_2008} found that appropriately scaled virtual avatars offer participants a recognizable size reference and a sense of connection to virtual objects and surroundings, which is also supported by works in~\cite{ogawa_virtual_2019} and~\cite{jung_over_2018}. Additionally, virtual self-avatars can improve participants’ distance perception~\cite{ebrahimi_investigating_2018, gonzalez-franco_individual_2019}. Thus, given that the awe experience is associated with an individual’s perception of the vastness of sublime stimuli~\cite{keltner_approaching_2003}, we hypothesized that altering an individual’s perception of objects’ size and distance may influence their sense of awe. To enhance embodiment, Kokkinara and Slater~\cite{kokkinara_measuring_2014} suggested visuomotor and visuotactile synchronous stimulation, and Bourdin et al.~\cite{bourdin_virtual_2017} proposed first-person virtual bodies. These approaches inspired our embodied design.

\subsection{VR Awe-Inspiring Experience Design}

In exploring VR design strategies for awe experiences, some scholars have turned to real-life awe-inspiring scenes. Quesnel et al.~\cite{quesnel_are_2018} immersed participants in four real panoramic videos: the Grand Canyon, snowy mountains, urban landscapes, and cosmic scenes. They found that factors such as aesthetics, scale, familiarity, and personalization influence participants' sense of awe. However, they did not delve deeper into identifying the specific elements that induce awe. Chirico et al.~\cite{chirico_designing_2018} used the two characteristics of awe—perceived vastness and the need for accommodation~\cite{keltner_approaching_2003}—to create three VR scenes depicting awe-inspiring natural environments: forest scenes, snowy mountain vistas, and the Overview Effect. Results showed higher awe in these VR scenes than controls. However, they did not introduce new awe-inducing scenes that had not been previously studied. Thus, their design strategies may not be applicable to create new awe-inspiring scenes.

Some scholars have drawn inspiration from real-life experiences. Kitson and Riecke~\cite{kitson_can_2018, kitson_are_2018} focused on lucid dreams, while Quesnel~\cite{quesnel_creating_2018} and Stepanova~\cite{stepanova_understanding_2019,stepanova_spacevirtual_2019} simulated the Overview Effect in VR to evoke awe. However, their design elements were primarily tailored for lucid dream and Overview Effect experiences in VR, limiting their applicability to other awe experiences. In subsequent studies, Liu et al.~\cite{liu_virtual_2022} combined lucid dreams, the Overview Effect, and VR flight to create a new awe experience, they offered preliminary exploration into the role of embodiment in awe experience in VR.

In summary, existing research primarily analyzes awe-inspiring natural scenes and individual experiences without providing a widely applicable design rationale. Therefore, we aimed to systematically derive design elements from comprehensive theories related to the sublime, which detail the characteristics of emotion elicitors.

\section{Sublime-based Awe-inspiring VR Design Elements}

We systematically extracted several sublime concepts from philosophical literature and organized them into three high-level concepts: terror, distance or modification, and obscurity. Philosophers have provided further explanations and examples for these three concepts in their theories. We organized them into more specific middle-level concepts to guide design practice (see \cref{tab1}).

\subsection{Terror}

Terror can be achieved through power and magnitude~\cite{burke_philosophical_2019}.  Objects of power can originate from either nature or authority. They can demonstrate their power by exhibiting superior dynamical power compared to significant obstacles~\cite{burke_philosophical_2019}, such as delivering powerful signs such as creating excessive loudness or causing painful sounds from surrounding creatures. Objects of magnitude can display mathematical greatness beyond comparison, such as appearing extremely large or small~\cite{schopenhauer_world_1969,kant_critique_2009}. 

\subsection{Distance or Modification}

Distance or modification can be achieved by making individuals realize that they are safe, such as when sublime objects do not contain the intent of attack, or when individuals are finally liberated from danger. It can also be achieved by making the sublime objects understandable for individuals. For instance, the immersive objects should not be too confusing, such as appearing monstrous or colossal, for individuals to apprehend their greatness~\cite{burke_philosophical_2019,kant_critique_2009}.

\subsection{Obscurity}

Obscurity can be achieved by altering the shape, color, etc., of objects to induce confusion and uncertainty~\cite{burke_philosophical_2019}. Visual elements such as darkness or extreme brightness, which differ from the immediate surroundings, can make an object striking. Colors like black and brown can intensify the obscure situation. Objects’ modes of movement, such as suddenness, intermitting, and quick transition, can also convey uncertainty.

\begin{table}[tb]
  \caption{\centering High-level concepts, middle-level concepts, and VR design elements in VR scenes.}
  \label{tab1}
  \scriptsize%
  \centering%
  \begin{tabu} to \columnwidth {%
    X[1,l]  
    X[2,l]
    X[2,l]
    }
    \toprule
    High-level concepts & Middle-level concepts & Design elements in VR scenes \\
    \midrule
    Terror & Power (natural power, the power that arises from authority), excessive loudness, painful sounds, magnitude (objects of extremely large or small, infinity)~\cite{burke_philosophical_2019,schopenhauer_world_1969,schopenhauer_world_1969}. & Huge waves rise from the sea, accompanied by a tremendous sound; a vast moon suddenly appears overhead, casting extreme light in contrast to the darkness of the night. \\
    Distance or modification & Safety, understandability ~\cite{burke_philosophical_2019,schopenhauer_world_1969,kant_critique_2009}. & The huge waves are at a distance from the shore, and after rising, they gradually move away. The vast moon eventually sets, as the sun begins to rise. \\
    Obscurity & Unfamiliar situations, extreme darkness or light, sad and fuscous colors (brown, black), suddenness, intermitting, quick transition ~\cite{burke_philosophical_2019,schopenhauer_world_1969,kant_critique_2009}. & The vast moon is much larger than what we usually see in daily life. \\
    \bottomrule
  \end{tabu}
\end{table}

\section{The VR Experience Design}

\begin{figure}[tb]
 \centering 
 \includegraphics[width=\columnwidth]{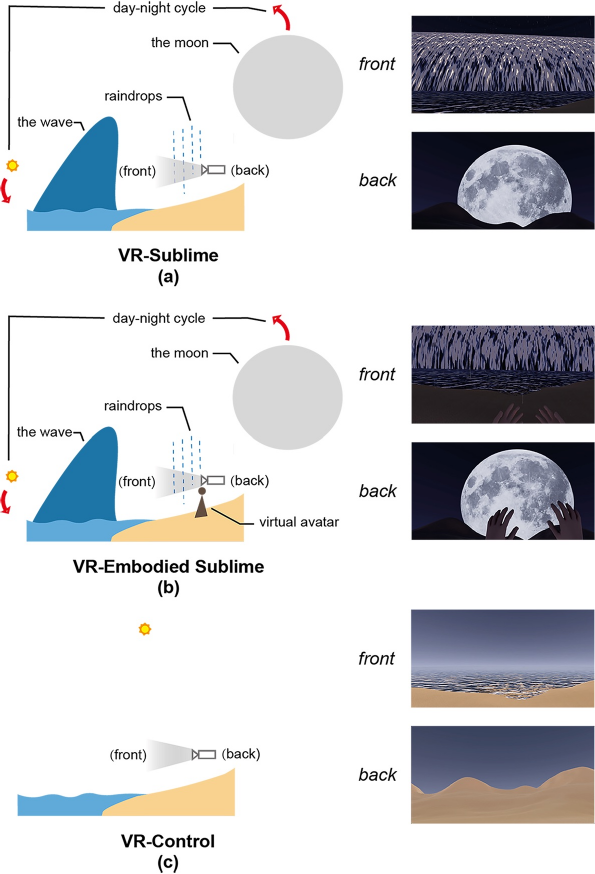}
 \caption{Three versions of the VR design: (a) VR-Sublime, (b) VR-Embodied Sublime, and (c) VR-Control.}
 \label{fig1}
\end{figure}

In order to understand the effectiveness of sublime and embodied design elements in inducing awe, we developed three VR scenes: \textbf{(1) VR-Sublime scene} developed based on sublime design elements; \textbf{(2) VR-Embodied Sublime scene}, including embodied experience based on VR-Sublime design; \textbf{(3) VR-Control scene}, excluding sublime and embodied design elements. The following subsections describe in detail how environments were created (see \cref{fig1}). 

The VR-Sublime scene was designed based on sublime-based awe-inspiring design elements around the theme of tidal forces, Table 1 illustrates how three sublime design elements can be used to create awe-inspiring scenes through specific examples. However, other themes, such as towering buildings or powerful tornadoes, can also utilize these design elements. In our case, the magnitude of the vast moon and huge waves, the tremendous sound, and the extreme moonlight in contrast to the darkness of night conveyed terror. The obscurity was achieved by making the vast moon much larger than what we usually see in daily life. The huge waves were at a distance from the shore and moved away after rising, and the moon eventually set, which conveys a sense of safety.

\begin{figure*}[tb]
 \centering 
 \includegraphics[width=\textwidth]{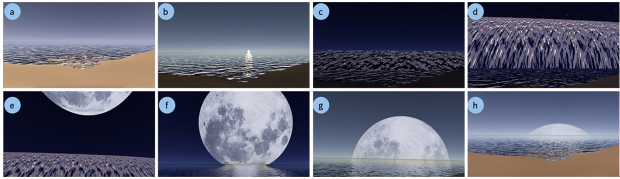}
 \caption{(a) A viewer finds himself/herself standing on the beach with a virtual body. (b) The viewer witnesses the gradual descent of the sun. (c-d) After the sunset, a huge wave begins to rise from the sea at a distance from the viewer. (e) When the waves reach halfway to their highest point, virtual rain falls. Meanwhile, a vast moon slowly ascends since sunset, eventually reaching the upper part of the viewer's field of vision. (f-g) The huge wave gradually moves away from the viewer and descends back into the sea, with the moon also slowly descending. (h) After the moon sets, the sun rises again, casting warm sunlight on the beach.}
 \label{fig2}
\end{figure*}

The entire VR experience includes the following steps (see \cref{fig2}). Initially, the viewer finds themselves standing on the beach surrounded by the sea near sunset. Next, the viewer witnesses the gradual descent of the sun. After the sunset, a huge wave begins to rise from the sea at a distance from the viewer. When the waves reach halfway to their highest point, virtual rain falls. Meanwhile, a vast moon slowly ascends since sunset, eventually reaching the upper part of the viewer's field of vision. As the viewer continues to observe the rising waves, they may perceive the vast moon suddenly appearing in the sky. Subsequently, the huge wave gradually moves away from the viewer and descends back into the sea, with the moon also slowly descending. Lastly, after the moon sets, the sun rises again, casting warm sunlight on the beach. The entire experience concludes as the sun rises overhead. 

\begin{figure}[tb]
 \centering 
 \includegraphics[width=\columnwidth]{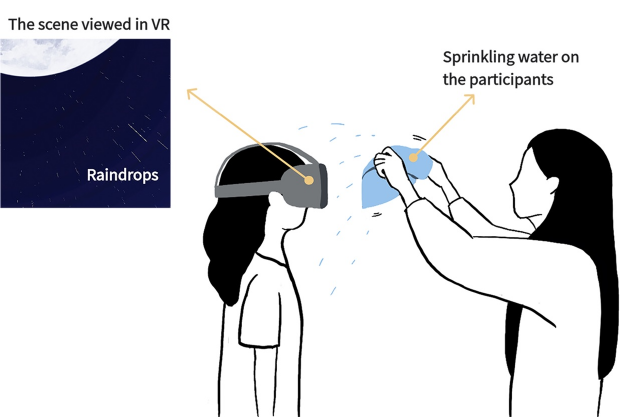}
 \caption{The research facilitator sprinkled water on the participant while she was experiencing “raindrops” in the VR environment.}
 \label{fig3}
\end{figure}

The VR-Embodied Sublime scene adds a body ownership illusion based on the VR-Sublime scene, achieved by utilizing visuomotor and visuotactile synchronous stimulation~\cite{kokkinara_measuring_2014}. The viewers have a complete virtual body in the scene. Visuomotor synchronization involves coordinating the viewer's movements with the virtual body's actions, enabling control over virtual arms and head rotation for a 360-degree view. The viewer can stand in place and turn around to observe the elements within the scene, but they cannot interact with their surroundings. The virtual body matches the viewer's gender, aiding in assessing body size. Visuotactile synchronization involves sprinkling water on the viewer to simulate "raindrops" when waves reach halfway (see \cref{fig3}).

While this work is conceptually similar to that of ~\cite{chirico_designing_2018}, our approach is based on sublime design elements rather than specific awe-inspiring content. This method allows us to vary our content rationally and flexibly according to three sublime design elements, creating new scenes and preventing the decrease in the intensity of awe from repeated exposure to the same content. For example, content that satisfies the element of terror might include intimidating entities such as superheroes with extraordinary abilities or massive artificial constructs. Obscurity elements could involve observing the slow-motion blending of paint or exploring the microscopic world—experiences rarely encountered in daily life. Element of distance or modification might include surrounding participants with a glass enclosure to create a sense of safety from ocean waves. Additionally, beyond awe-inspiring elements, we designed an experience incorporating embodied design elements to explore the effects of embodied experience. This approach expands our understanding of VR's impact on awe experiences.

The VR-Control scene only shows a static seaside scene without the sublime and embodied design elements.

\section{Methods}

\begin{figure}[tb]
 \centering 
 \includegraphics[width=\columnwidth]{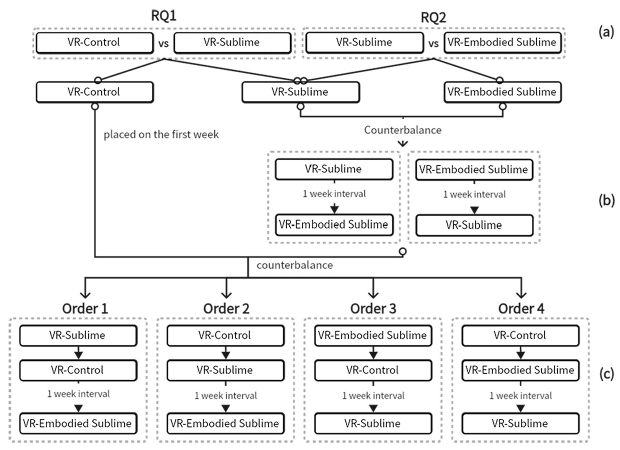}
 \caption{Experimental design for research questions.}
 \label{fig4}
\end{figure}

Since individual differences may lead to varying perceptions of awe, such as openness to experience~\cite{silvia_openness_2015}, we used a within-subject experimental design to mitigate the individual differences. We conducted a two-week within-subject experiment with 28 participants (ages 18-26) in a university lab~\cite{liu_virtual_2022}. The aim was to investigate the effectiveness of various VR experience designs in invoking awe experiences and identifying potential contributing design features. Each participant used all three VR designs, and the order was counterbalanced. RQ1 was addressed by comparing the results between the VR-Control scene and the VR-Sublime scene to evaluate the effectiveness of sublime design elements in invoking awe. RQ2 was addressed by comparing the results between the VR-Sublime scene and the VR-Embodied Sublime scene to evaluate the effectiveness of embodied design elements in impacting the awe experience (see \cref{fig4}-a). 

To minimize potential carry-over effects~\cite{quesnel_are_2018} (i.e., the viewer may feel a lower intensity of emotion when they experience a similar scene again within a short period), we designed a 1-week interval between the experience of VR-Sublime scene and VR-Embodied sublime scene (see \cref{fig4}-b). Since viewers typically have more impressive and richer feelings when they first experience a scene, we placed the VR-Control scene in the first week, allowing them to compare it with the other two scenes. Thus, after counterbalancing, there are 4 orders (see \cref{fig4}-c).

\begin{figure*}[tb]
 \centering 
 \includegraphics[width=\textwidth]{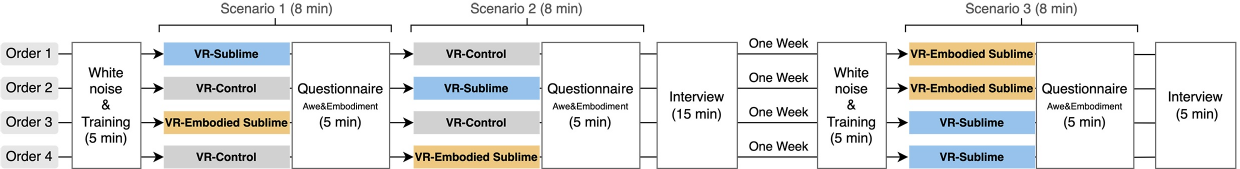}
 \caption{Study procedure.}
 \label{fig5}
\end{figure*}

\subsection{Participants}

We recruited 28 participants (P1-P28) aged 18-26 (M = 20.5, SD = 2.67), comprising 15 females and 13 males. Participants were recruited through a public release channel of our university's mental health center. All participants were undergraduate or graduate students with diverse majors, including Chinese Language and Literature, Data Science, Advertising, and Media Music. Their previous experience with VR technology varied: 4 participants had never used VR, 21 had used it only occasionally, and 3 used it almost monthly. None reported a history of seizures, severe headaches, or uncorrected eye or ear conditions. All participants completed the consent forms of the study, which were approved by our university’s ethical board. We compensated participants with 50CNY after completing all tests.

\subsection{Procedure}

The entire study was conducted with each participant individually in a quiet room at our university's lab, facilitated by two researchers. Before the study, each participant completed a demographic questionnaire and received a brief introduction to the study process. Participants then randomly experienced three scenes in one of the four orders (see \cref{fig5}). Each participant underwent the three VR scenes (i.e., VR-Control, VR-Sublime, and VR-Embodied Sublime). 

Before each session, participants listened to 2-minute white noise to help them relax and then completed a training session to become familiar with VR usage. Each VR scene lasted approximately 3 minutes. After experiencing each VR scene, participants completed two questionnaires on awe and embodiment, each lasting 5 minutes. After completing the questionnaires of the second VR scene in week 1 and the third VR scene in week 2, they participated in an approximately 15-minute semi-structured interview. The researcher facilitators asked participants several open-ended questions about their emotional responses and the influences of sublime-related and embodied design elements on their feelings (e.g., ``\textit{How did you feel when you were observing the rising wave?}"). The entire process for each participant lasted approximately 60 minutes. Participants could stop whenever they feel uncomfortable. 

\subsection{Data Collection and Analysis}

To address RQ1, we utilized the Awe Experience Scale (AWE-S-6) to assess participants' awe experiences~\cite{yaden_development_2019}, and utilized R version 4.4.1~\cite{r_core_team_r_2013} and lme4 version 1.1.35.5~\cite{bates_lme4_2007} to conduct linear mixed effects analyses of awe ratings. To determine if there was a statistically significant mean difference in awe ratings between VR-Control scene and VR-Sublime scene, we compared models with and without sublime design elements through likelihood ratio tests. We conducted semi-structured interviews to help explain the quantitative results and explore the influence of the sublime design elements on the awe experience. All interviews were recorded using a video camera, and thematic analysis~\cite{braun_using_2006} was utilized to analyze interview data by three research facilitators.

To address RQ2, we utilized the AWE-S-6 to assess participants' awe experiences~\cite{yaden_development_2019}, and the VR Avatar Embodiment Questionnaire to evaluate their embodiment scores~\cite{peck_avatar_2021}. To determine if there was a statistically significant mean difference in embodiment rating and awe ratings between VR-Sublime scene and VR-Embodied Sublime scene, we also conducted linear mixed effects analyses to compare models with and without embodied design elements both in awe ratings and embodied ratings through likelihood ratio tests. The semi-structured interviews also include questions aimed at understanding the quantitative results and the influence of embodied design elements. 

We also compared the data of VR-Control and VR-Embodied Sublime to present the effect of the VR-Embodied Sublime scene alone. To determine if there was a statistically significant difference in mean embodiment and awe ratings between the two scenes, we conducted linear mixed effects analyses. We compared models with and without both sublime and embodied design elements in both awe and embodiment ratings through likelihood ratio tests.

We assessed the impact of order and novelty on all measures, aiming to eliminate their potential influence on awe ratings. Order effects were examined using a mixed 2 (sublime/embodied design elements: without vs. with) × 4 (order: 1st vs. 2nd vs. 3rd vs. 4th) ANOVA~\cite{liu_virtual_2022}. The consideration of novelty arises from the possibility that participants new to VR might experience novelty, which may also induce feelings of awe to some extent~\cite{keltner_approaching_2003}. To address this potential issue, we conducted a mixed 2 (sublime/embodied design elements: without vs. with) × 3 (previous VR experience) ANOVA. Cohen’s \textit{d} was used to measure effect size, and the data were adjusted for random effects by subtracting the estimated random effects for each subject from their corresponding awe ratings.

We used thematic analysis~\cite{braun_using_2006} to analyze participants’ responses to open-ended questions about their experience towards the sublime-related and embodied design elements. We transcribed 784 minutes of recorded audio data into texts. Initially, three authors familiarized and coded 14\% of the materials independently, and then discussed together to generate the preliminary codes and potential themes. Subsequently, the first author coded the remaining materials, while the other two authors independently coded half of the remaining data. The three authors engaged in multiple discussions of codes and data to produce collaborative interpretations, progressing from open coding to theme discussions. Disagreements were resolved through successive rounds of synchronous review. Ultimately, the three authors collaboratively identified several consistent themes, which will be elaborated in the results section.

\section{Results}

The results showed that the sublime design elements significantly elicited awe experiences, while the addition of embodied elements alongside sublime ones did not further enhance the intensity of awe. Visual inspection of residual plots from all models did not reveal any obvious deviations from homoscedasticity or normality. We present our quantitative findings supplemented with qualitative data.

\subsection{Impact of Sublime-Related Design Elements on Awe Experience}

\subsubsection{Quantitative Results}

In the comparison of awe ratings between the VR-Control and VR-Sublime scenes. No significant order effects were found (all \textit{p} $>$ 0.14). However, a significant interaction between previous VR experience and sublime design elements was observed in explaining awe ratings, F (2, 25) = 3.68, \textit{p} $<$ 0.05. Post-hoc tests showed that the simple main effect of VR experience was not significant, both without and with sublime design elements (all \textit{p} $>$ 0.35), as indicated by Bonferroni-adjusted \textit{p}-values. Pairwise comparisons indicated no significant differences in mean awe ratings among different VR experience levels, under conditions of with or without sublime design elements (all \textit{p} $>$ 0.24). This suggests awe ratings were likely unaffected by users’ VR experience.

\begin{figure}[h]
 \centering 
 \includegraphics[width=0.5\columnwidth]{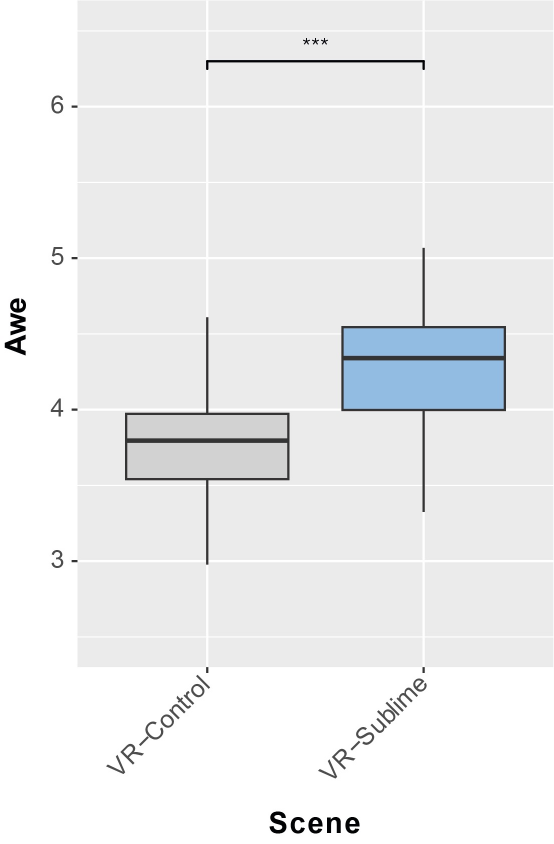}
 \caption{Awe ratings of VR-Control (left) and VR-Sublime (right).}
 \label{fig6}
\end{figure}

We conducted a linear mixed effects analysis to assess the impact of sublime design elements on awe ratings. Sublime elements were treated as fixed effects, with subject intercepts as random effects. Results revealed a significant effect of sublime design elements on awe ratings ($\chi^2(1)$ = 12.00, \textit{p} $<$ 0.05), rising it by approximately 0.56 ± 0.14 (standard errors). The effect size was 1.25 with a 95\% CI of [0.66, 1.83]. These findings showed that sublime design elements can evoke an awe experience (see \cref{fig6}). 

\subsubsection{Qualitative Results}

The qualitative findings were consistent with the quantitative results, with many people reporting a general post-experience emotion after experiencing VR scenes with sublime design elements, such as feeling “\textit{self-diminished}” (P5 and P15), “\textit{peaceful}” (P3, P4, P5, P6, P7, P10), and “\textit{a desire to experience again}” (P3). P5 expressed, “\textit{After experiencing those crazy waves, and then seeing this amazing vast moon, I was just soaking in this beautiful sight that you don't see every day, feeling really happy about it}”. These findings suggest that VR scenes with sublime design elements can effectively induce awe and positively influence participants' emotions. In addition to the general experience, participants reported how different sublime design elements impacted their feelings.

\textbf{\textit{Terror arose from the inherent power, relative size and distance, and uncertainty of objects.}} Participants perceived the huge waves as inherently powerful, as waves are often metaphors for destructive power. For example, P2 commented that the waves \textit{“might engulf people, evoking a sense of natural peril}”, and P5 reported that she had “\textit{once dreamt of a tidal wave approaching home}”. The waves also implied the great force of the moon, with nearly half of the participants recognizing the tidal effects of the moon. For instance, P3 said, “\textit{I realized the seawater in front was attracted by the moon, indicating tides, which were then gradually pushed back by the moon. This process is quite startling.}” and P19 mentioned that “\textit{This natural scene underwent powerful changes, which were awe-inspiring.}” Even though the wave was inherently powerful, only when it had a relatively large size and close distance to the participant can it invoke their sense of terror.  For instance, P1 stated, “\textit{I initially thought it was a small wave, but as it grew taller, reaching my height, it became frightening,}” and P6 reported, "\textit{When the wave rose up, I felt it was too close to me, a bit frightening.}" Uncertainty of the threatening objects could keep participants stay in the feeling of nervous and worried, for instance, "\textit{The second time I experienced it, I knew the sea level would rise, so I wasn't too scared. But I didn't know I would feel the raindrops, which made me a bit more afraid}" (P14) and “\textit{When I first saw the moon, I was hesitant to move my body much because I wasn't sure if it was going to fall or just remain in its position. There was a momentary unease in my heart. But once it descended normally, that feeling of tension disappeared}” (P19).

\textbf{\textit{Obscurity arose from the rarity of elements and can be reasonably explained.}} The rare elements can manifest as the unusual presence of ordinary things or uncommon experiences. When the vastness of the moon appeared, most participants found it difficult to comprehend because it deviated from the typical appearance of the moon in real life. For instance, P9 stated, “\textit{At first, the connection with reality was strong, starting with the simple beach and sea, and then suddenly there appeared a very abnormal moon.}” P11 mentioned, “\textit{This feeling arose because it did not align with the real world.}” P14 said, “\textit{Under those circumstances, the moon felt closer to where we stood, which was shocking as it completely deviated from real-world experience.}” However, P9 and P19 did not perceive the vast moon as unusual because they had seen many pictures of it before. In contrast, P19 found the huge wave more startling because he had only visited the sea once. Similarly, during the second week when participants observed the waves and moon again, most of them reported reduced intensity of awe due to familiarity, for instance, “\textit{Because I had already experienced it once before, so this time, my emotions didn't fluctuate as much}” (P15). Subsequently, many participants realized the connection between the tidal gravity of the vast moon and the emergence of the huge wave. For instance, “\textit{I began to wonder why the wave was so massive, but upon seeing the moon, my question was resolved—it was due to the tidal effect}” (P10). “\textit{After linking the moon with the rising waves, I better recognized the connection between the two things}” (P13). The unusual event could be reasonably explained, which deeply resonated with the participants, for instance, “\textit{Because the attack of that wave could potentially happen in real life, so the threat and fear felt very real}” (P1) and “\textit{And because there was a correlation between the huge waves and the moon, it triggered more associations, which made it more impactful}” (P9). Besides, the obscurity should be linked to the objects of terror, for instance, P19 highlighted “\textit{It's not just any unusual event that evokes that kind of awe; for example, if a person suddenly disappears, even though it's not real, it won't necessarily leave you in awe. It's mainly because this unusual event is related to the immense power of nature that it brings about that sense of awe.}”

\textbf{\textit{Distance or modification arose from the anticipation that threatening objects would not cause substantial harm.}} The alleviation of fear is essential for the emergence of awe; otherwise, participants may be overwhelmed by negative emotions, hindering their experience of awe. Participants felt their fear was alleviated when the terrible objects displayed no harm to them, such as stopping rising (P12, P20, P22, P26), stopping approaching (P4, P6, P11), and completely disappearing (P1 and P14). For example, P14 described, “\textit{At first, I was concerned it might keep growing taller and eventually engulf me. But I realized it posed no threat to me. Instead, and it gradually receded, my fear diminished.}” Besides, it is the signs that the threat will not show up again that make participants feel completely relieved, for instance, P7 mentioned, “\textit{In this transition from day to night, you could see a pattern that you recognized, so when the sun rose again, I felt relieved because it had become normal,}” and P8 stated, “\textit{I felt completely calm because, after the sun rose, it returned to the initial scene I had entered.}” The virtual nature of VR inherently introduces some degree of distance and modification, for instance, “\textit{Given VR's limitations, it's unlikely to fully replicate the authentic sensation of being in a real ocean, such as feeling the sea breeze. So, I knew it wasn't real from the start}” (P2) and “\textit{In reality, witnessing such a massive tsunami would terrify me. But in this virtual world, I felt safe, so I wasn't as scared}” (P6). As a result, some participants identified the physical sensation of a 'raindrop' as artificial, despite its realistic feel, because they were certain it couldn't occur in VR. For instance, P1 mentioned that “\textit{Although the raindrops gave a physical sensation, knowing they were artificial made me more aware that I wasn't in a real environment, reducing my fear of the waves,}” and P9 explained that “\textit{Despite the correlation between rain and the scene, and the realistic sensation in my body, I still believed it wasn't raining in VR, but rather in the real world.}”

\subsection{Impact of Embodied Design Elements on Awe Experience}

\subsubsection{Quantitative Results}

In the comparison of awe ratings between the VR-Sublime and VR-Embodied Sublime conditions. No significant novelty effects were found (all \textit{p} $>$ 0.11). However, a significant interaction between test order and embodied design elements was observed in explaining awe ratings, F (3, 24) = 5.90, \textit{p} $<$ 0.05. Post-hoc tests showed that the simple main effect of the VR experience was not significant, both without and with sublime design elements (all \textit{p} $>$ 0.27). Pairwise comparisons indicated no significant differences in mean awe ratings among different VR experience levels, under conditions of with or without sublime design elements (all \textit{p} $>$ 0.15). This suggests that awe ratings were likely unaffected by order effects. 

\begin{figure}[h]
 \centering 
 \includegraphics[width=\columnwidth]{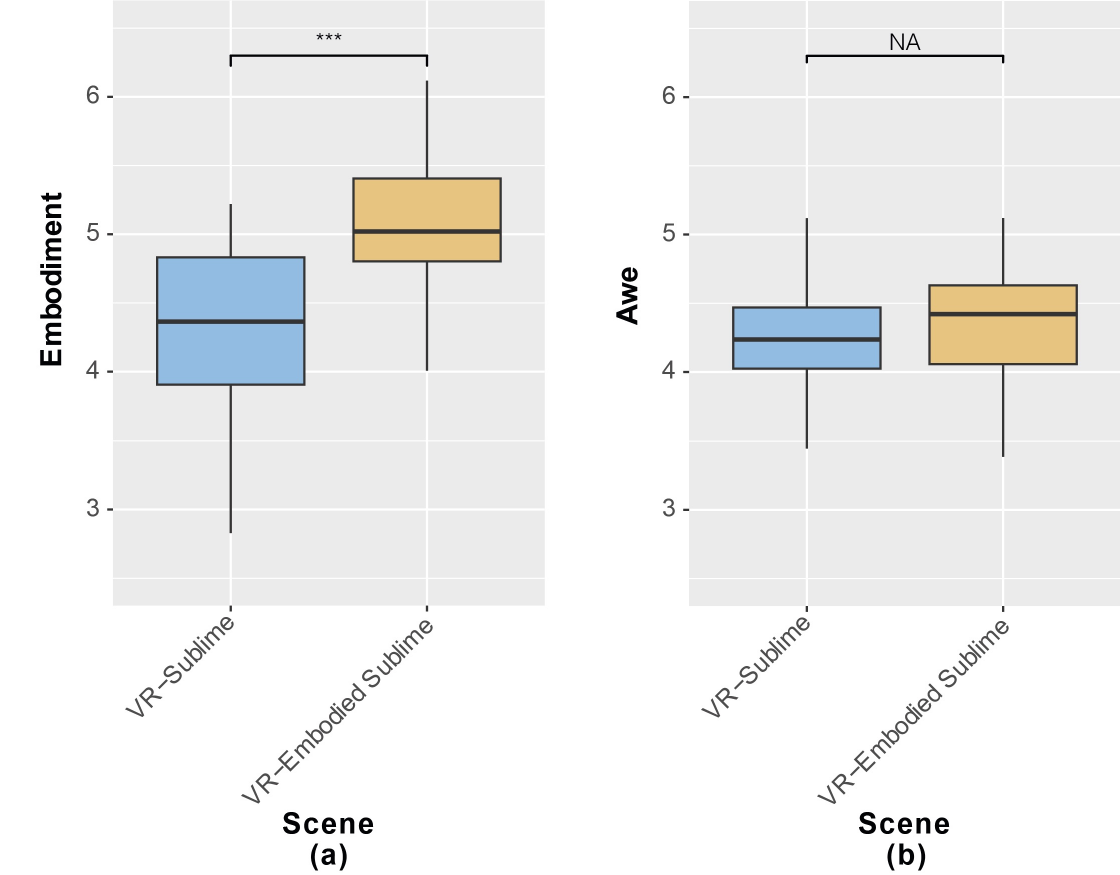}
 \caption{(a) Embodiment ratings and (b) Awe ratings of VR-Sublime (left) and VR-Embodied Sublime (right).}
 \label{fig7}
\end{figure}

We conducted a linear mixed effects analysis to assess the impact of embodied design elements on embodiment ratings and awe ratings. Embodied design elements were treated as fixed effects, with subject intercepts as random effects. Results revealed a significant effect of embodied design elements on embodiment ratings ($\chi^2(1)$ = 14.68, \textit{p} $<$ 0.05), raising it by approximately 0.83 ± 0.19 (standard errors). The effect size was 1.28 with a 95\% CI of [0.69, 1.87]. These findings indicated that embodied design elements did enhance participants' sense of embodiment. However, there was no statistically significant difference in the awe ratings between VR-Sublime and VR-Embodied Sublime ($\chi^2(1)$ = 0.47, \textit{p} = 0.49). The results suggested that VR design with embodied design elements did not lead to higher awe compared to the non-embodied one (see \cref{fig7}).

\subsubsection{Qualitative Results}

\textbf{\textit{Embodied experience enhanced immersion but had minimal impact on emotion.}} The absence of the virtual avatar made many participants (P5, P6, P9, P12, P14, P19-P22) feel like they were watching a film rather than being physically situated within it. Conversely, when the virtual avatar was present, participants perceived themselves as engaged within the virtual world, for instance, “\textit{The raindrops were so immersive that I felt completely within the scene, almost as if I had merged into a virtual body}” (P5) and “\textit{When I lifted my hand, it moved too, heightening my sense of reality}” (P7). They also expressed a desire to interact with virtual objects when they have a virtual avatar, such as “\textit{approach the sea and touch the water}” (P19, P6, P10) or “\textit{grasp the sand}” (P6, P20). However, the embodied experience did not influence their emotion, as P2 mentioned “H\textit{aving a body affects immersion, but not emotions.}”

\textbf{\textit{Participants felt body presence even in the absence of virtual avatars.}} Even without virtual avatars, participants can still imagine themselves having a body within the virtual environment, for instance, “\textit{Because I couldn't see my own body, as long as I didn’t look down, I felt like my entire body existed at that moment.}” (P13) and “\textit{If there was no virtual body, my immersion into the virtual world might be slow at first, but gradually, I started imagining a body, leading to a smoother experience}” (P19). And they felt like they could be attacked by the huge waves as stated in Section 6.1.2. Without a virtual body, participants felt more freedom, for instance, “\textit{Without that virtual avatar, I wouldn't feel constrained in my movements}” (P1) and “\textit{Without a body, I found my imagination expanded. For example, with a body, I can't touch the water surface, but without one, I can imagine touching it.} (P19).”

\textbf{\textit{The virtual body influenced participants’ perceptions of size and distance, but not much.}} The virtual avatar provided participants with a frame of reference, enabling them to determine their position and quantify the distance of surrounding objects (P2, P5, P6, P12, P14, P15), as well as allowing them to compare the size between their virtual avatar and the sublime objects (P7, P12, P23). For example, P23 mentioned, “\textit{I compared the size of my hand with the moon. It felt like a grand spectacle that's hard to imagine. In front of this moon, I felt extremely insignificant.}” The presence or absence of the virtual avatar did influence some participants' perception of the size and distance of the vast moon and huge wave (P2, P5, P10), although the effect was minimal, as P7 commented, “\textit{Comparing the size of objects in this scene with the size of my character, it felt about right. It didn't exceed my usual perception.}”

\subsection{The Comparison between VR-Control Scene and VR-Embodied Sublime Scene}

In the comparison of awe ratings between the VR-Control and VR-Embodied Sublime scenes. No significant order effects (all \textit{p} $>$ 0.11) or novelty effects (all \textit{p} $>$ 0.61) were found. 

We conducted a linear mixed effects analysis to investigate the difference between VR-Control and VR-Embodied Sublime. Sublime and embodied design elements together were treated as fixed effects, with subject intercepts as random effects. Results revealed a significant effect of sublime and embodied design elements together on embodiment ratings ($\chi^2(1)$ = 20.23, \textit{p} $<$ 0.05), raising it by approximately 1.07 ± 0.20 (standard errors). The effect size was 1.68 with a 95\% CI of [1.06, 2.31]. These findings indicated that sublime and embodied design elements together enhanced participants' sense of embodiment. 

\begin{figure}[h]
 \centering 
 \includegraphics[width=\columnwidth]{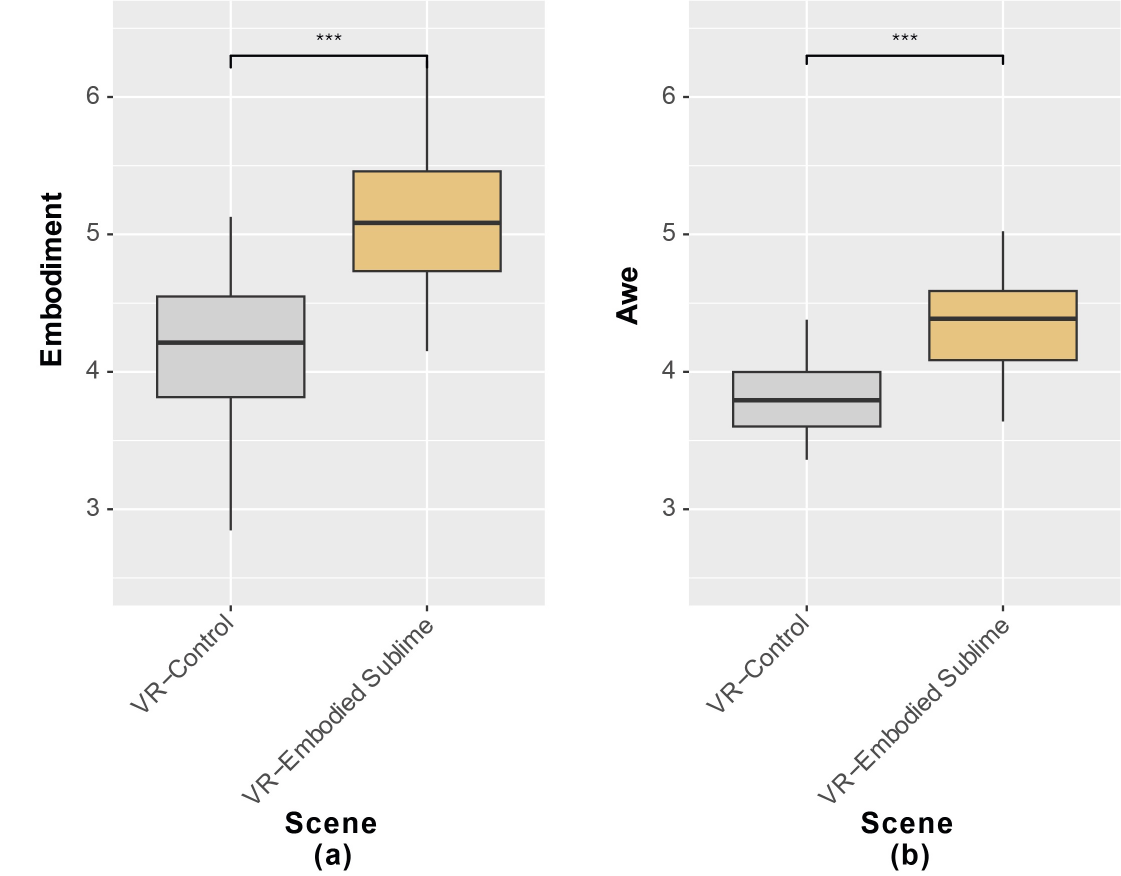}
 \caption{(a) Embodiment ratings and (b) Awe ratings of VR-Control (left) and VR-Embodied Sublime (right).}
 \label{fig8}
\end{figure}

Results also revealed a significant effect of sublime and embodied design elements together on awe ratings ($\chi^2(1)$ = 17.48, \textit{p} $<$ 0.05), raising it by approximately 0.66 ± 0.13 (standard errors). The effect size was 1.65 with a 95\% CI of [1.03, 2.27]. These findings showed that sublime and embodied design elements together evoked an awe experience (see \cref{fig8}). 

\section{Discussion}

\subsection{Threat-Based Awe Experience Can Lead to Positive Results if Modified by Reasonable Events and Elements with Positive Metaphors}

As hypothesized (RQ1), design elements associated with the sublime evoked awe in participants. Contrary to concerns raised
by Monroy and Keltner~\cite{monroy_awe_2022}, our threat-based awe-inspiring experience did not elicit negative emotional responses among most participants; instead, participants reported feeling peaceful and experiencing a sense of pleasure. There could be two reasons for this. Firstly, the event was reasonable for participants because there was a connection between the huge waves, the vast moon, and the rain. As discussed in Section 6.1.2, many participants understood that the huge waves were caused by the tidal effect of the vast moon. Therefore, participants' needs for accommodation were satisfied, leading to feelings of enlightenment~\cite{keltner_approaching_2003}. Secondly, many participants found the view of the vast moonset to be splendid and touching. This could be attributed to the influence of Chinese culture's preference for the moon~\cite{monroy_awe_2022}, thereby enhancing the overall positive awe experience~\cite{keltner_approaching_2003}.

Kant appreciated danger because it can uplift the mindset of those who face it, especially when they show courage in challenging situations~\cite{kant_critique_2009}. We agree after witnessing the joy and excitement reported by many participants after experiencing a terrifying event. Therefore, we recommend that future research explore ways to adjust threat-based awe experiences to maintain their beneficial effects while eliciting positive emotional responses.

\subsection{Trade-offs in Design Elements}

Our results indicate that there are some trade-offs between design elements. There needs to be a balance between terror and distance or modification. If the level of terror is too high, it may make participants feel scared and unable to fully experience awe, while if the level of distance or modification is too high, it may be difficult to evoke intense emotions in participants. When designing awe-inspiring experiences using VR, it is important to consider that the virtual nature of VR itself is a form of distance or modification, which, to some extent, makes participants feel safe. Additionally, there needs to be a balance between obscurity and reasonableness. If things are too obscure, participants may feel unable to understand and become frustrated, while if things are too easily understood, it may not invoke a need for accommodation in participants. For example, two participants felt that the vast moon was not surprising because they were familiar with it from pictures in daily life.

\subsection{Are Virtual Bodies Worth Considering in the Design of VR Awe Experience?}

Contrary to our expectations (RQ2), while successfully invoking a sense of embodiment, the embodied experience did not enhance the intensity of the awe experience. In our study, there are three possible explanations for this result. Firstly, due to the sense of presence, participants’ minds and bodies reacted as if terrifying events were occurring around them even without virtual bodies~\cite{freeman_virtual_2017}. Additionally, the awe experience shifted participants' focus from the self to the external world, suggesting that the virtual body may play a limited role in the awe experience. Secondly, as indicated by participants, we suppose that the size of the virtual body resembled the participants’ imagined body in scenes without embodied design elements. This similarity did not substantially alter their perception of size and distance, thereby having limited influence on their emotions. Thirdly, participants were not able to use their virtual body to move or interact with surrounding objects, which may have caused them to overlook the presence of the virtual body during the experience, resulting in a similar effect as when no virtual body was present~\cite{kim_active_2023}.

\subsection{Limitations}

We had three main limitations in the VR design and study design: (1) Our design of the virtual avatar did not adequately influence participants' perception regarding size comparisons and distances between their perceived bodies and surrounding objects, and also did not provide them with some interaction that is different from that without virtual bodies, future work can create a more distinctive embodied experience to explore its effect. (2) We only designed and tested one scenario based on sublime design elements, and more design practice is needed to fully validate the effectiveness of these sublime design elements in evoking awe. (3) Our subjects were all university students; future research should consider people from different age groups and backgrounds.

\section{Conclusion}

We designed an immersive VR experience grounded in theories of the sublime, enabling viewers to experience a day-night cycle with a vast moon and rising tide to evoke their sense of awe. Building upon this, we designed another VR experience with additional embodied design elements, allowing participants to view and control a virtual body and feel the sensation of raindrops. To understand the effectiveness of design elements in inducing awe, we conducted a within-subjects experiment with 28 participants who experienced both of our designed VR experiences, as well as a similar control design lacking both sublime and embodied elements. Our results showed that the VR experience with sublime design elements induced a significantly higher sense of awe compared to the one without, while the addition of embodied design elements did not enhance the intensity of the awe experience. To create an awe-inspiring VR experience, we suggest that (1) After implementing sublime design elements, enhance the overall positive emotional response by incorporating reasonable and beautiful elements. (2) Considering the trade-offs between terror and distance or modification to invoke adequate intense emotion, and the trade-offs between obscurity and reasonableness to invoke a need for accommodation. (3) Creating a more distinctive embodied experience for the viewer may influence their sense of awe.

\acknowledgments{
We thank the National Social Science and Arts Foundation (22BG137) for funding this project.}

\bibliographystyle{abbrv-doi}

\bibliography{paper}

\begin{thebibliography}{10}

\bibitem{arcangeli_awe_2020}
M.~Arcangeli, M.~Sperduti, A.~Jacquot, P.~Piolino, and J.~Dokic.
\newblock Awe and the {Experience} of the {Sublime}: {A} {Complex} {Relationship}.
\newblock {\em Frontiers in Psychology}, 11, 2020. doi: {{%
10\hspace{.1pt}\discretionary{.}{%
}{.}\hspace{.4pt}3389\discretionary{/}{%
}{/}fpsyg\hspace{.1pt}\discretionary{.}{%
}{.}\hspace{.4pt}2020\hspace{.1pt}\discretionary{.}{%
}{.}\hspace{.4pt}01340}}


\bibitem{bates_lme4_2007}
D.~M. Bates.
\newblock lme4: {Linear} mixed-effects models using {S4} classes.
\newblock {\em (No Title)}, 2007.

\bibitem{bourdin_virtual_2017}
P.~Bourdin, I.~Barberia, R.~Oliva, and M.~Slater.
\newblock A {Virtual} {Out}-of-{Body} {Experience} {Reduces} {Fear} of {Death}.
\newblock {\em PLOS ONE}, 12(1):e0169343, Jan. 2017.
\newblock Publisher: Public Library of Science. doi: {{%
10\hspace{.1pt}\discretionary{.}{%
}{.}\hspace{.4pt}1371\discretionary{/}{%
}{/}journal\hspace{.1pt}\discretionary{.}{%
}{.}\hspace{.4pt}pone\hspace{.1pt}\discretionary{.}{%
}{.}\hspace{.4pt}0169343}}


\bibitem{braun_using_2006}
V.~Braun and V.~Clarke.
\newblock Using thematic analysis in psychology.
\newblock {\em Qualitative Research in Psychology}, 3(2), Jan. 2006.
\newblock Publisher: Routledge \_eprint: https://www.tandfonline.com/doi/pdf/10.1191/1478088706qp063oa. doi: {{%
10\hspace{.1pt}\discretionary{.}{%
}{.}\hspace{.4pt}1191\discretionary{/}{%
}{/}1478088706qp063oa}}


\bibitem{burke_philosophical_2019}
E.~Burke.
\newblock {\em A {Philosophical} {Enquiry} into the {Origin} of {Our} {Ideas} of the {Sublime} and {Beautiful}}.
\newblock Columbia University Press, May 2019. doi: {{%
10\hspace{.1pt}\discretionary{.}{%
}{.}\hspace{.4pt}7312\discretionary{/}{%
}{/}burk90112}}


\bibitem{calvo_positive_2014}
R.~A. Calvo and D.~Peters.
\newblock {\em Positive {Computing}: {Technology} for {Wellbeing} and {Human} {Potential}}.
\newblock MIT Press, Nov. 2014.
\newblock Google-Books-ID: uI6ZBQAAQBAJ.

\bibitem{chen_how_2017}
W.~Chen, J.~Zhang, Y.~Qian, and Q.~Gao.
\newblock How disentangled sense of agency and sense of ownership can interact with different emotional events on stress feelings.
\newblock {\em Psicologia: Reflexão e Crítica}, 30:17, Sept. 2017.
\newblock Publisher: Curso de Pós-Graduação em Psicologia da Universidade Federal do Rio Grande do Sul. doi: {{%
10\hspace{.1pt}\discretionary{.}{%
}{.}\hspace{.4pt}1186\discretionary{/}{%
}{/}s41155\discretionary{%
}{-}{-}017\discretionary{%
}{-}{-}0071\discretionary{%
}{-}{-}y}}


\bibitem{chirico_designing_2018}
A.~Chirico, F.~Ferrise, L.~Cordella, and A.~Gaggioli.
\newblock Designing {Awe} in {Virtual} {Reality}: {An} {Experimental} {Study}.
\newblock {\em Frontiers in Psychology}, 8:2351, Jan. 2018. doi: {{%
10\hspace{.1pt}\discretionary{.}{%
}{.}\hspace{.4pt}3389\discretionary{/}{%
}{/}fpsyg\hspace{.1pt}\discretionary{.}{%
}{.}\hspace{.4pt}2017\hspace{.1pt}\discretionary{.}{%
}{.}\hspace{.4pt}02351}}


\bibitem{clewis_intersections_2022}
R.~R. Clewis, D.~B. Yaden, and A.~Chirico.
\newblock Intersections {Between} {Awe} and the {Sublime}: {A} {Preliminary} {Empirical} {Study}.
\newblock {\em Empirical Studies of the Arts}, 40(2):143--173, July 2022.
\newblock Publisher: SAGE Publications Inc. doi: {{%
10\hspace{.1pt}\discretionary{.}{%
}{.}\hspace{.4pt}1177\discretionary{/}{%
}{/}0276237421994694}}


\bibitem{ebrahimi_investigating_2018}
E.~Ebrahimi, L.~S. Hartman, A.~Robb, C.~C. Pagano, and S.~V. Babu.
\newblock Investigating the {Effects} of {Anthropomorphic} {Fidelity} of {Self}-{Avatars} on {Near} {Field} {Depth} {Perception} in {Immersive} {Virtual} {Environments}.
\newblock In {\em 2018 {IEEE} {Conference} on {Virtual} {Reality} and {3D} {User} {Interfaces} ({VR})}, pp. 1--8, Mar. 2018. doi: {{%
10\hspace{.1pt}\discretionary{.}{%
}{.}\hspace{.4pt}1109\discretionary{/}{%
}{/}VR\hspace{.1pt}\discretionary{.}{%
}{.}\hspace{.4pt}2018\hspace{.1pt}\discretionary{.}{%
}{.}\hspace{.4pt}8446539}}


\bibitem{freeman_virtual_2017}
D.~Freeman, S.~Reeve, A.~Robinson, A.~Ehlers, D.~Clark, B.~Spanlang, and M.~Slater.
\newblock Virtual reality in the assessment, understanding, and treatment of mental health disorders.
\newblock {\em Psychological Medicine}, 47(14):2393--2400, Oct. 2017.
\newblock Publisher: Cambridge University Press. doi: {{%
10\hspace{.1pt}\discretionary{.}{%
}{.}\hspace{.4pt}1017\discretionary{/}{%
}{/}S003329171700040X}}


\bibitem{gonzalez-franco_individual_2019}
M.~Gonzalez-Franco, P.~Abtahi, and A.~Steed.
\newblock Individual {Differences} in {Embodied} {Distance} {Estimation} in {Virtual} {Reality}.
\newblock In {\em 2019 {IEEE} {Conference} on {Virtual} {Reality} and {3D} {User} {Interfaces} ({VR})}, pp. 941--943, Mar. 2019.
\newblock ISSN: 2642-5254. doi: {{%
10\hspace{.1pt}\discretionary{.}{%
}{.}\hspace{.4pt}1109\discretionary{/}{%
}{/}VR\hspace{.1pt}\discretionary{.}{%
}{.}\hspace{.4pt}2019\hspace{.1pt}\discretionary{.}{%
}{.}\hspace{.4pt}8798348}}


\bibitem{gordon_dark_2017}
A.~M. Gordon, J.~E. Stellar, C.~L. Anderson, G.~D. McNeil, D.~Loew, and D.~Keltner.
\newblock The dark side of the sublime: {Distinguishing} a threat-based variant of awe.
\newblock {\em Journal of Personality and Social Psychology}, 113(2), 2017.
\newblock Place: US Publisher: American Psychological Association. doi: {{%
10\hspace{.1pt}\discretionary{.}{%
}{.}\hspace{.4pt}1037\discretionary{/}{%
}{/}pspp0000120}}


\bibitem{graves_friluftsliv_2020}
M.~Graves, H.~S. Løvoll, and K.-W. Sæther.
\newblock Friluftsliv: {Aesthetic} and {Psychological} {Experience} of {Wilderness} {Adventure}.
\newblock In M.~Fuller, D.~Evers, A.~Runehov, K.-W. Sæther, and B.~Michollet, eds., {\em Issues in {Science} and {Theology}: {Nature} – and {Beyond}: {Transcendence} and {Immanence} in {Science} and {Theology}}, Issues in {Science} and {Religion}: {Publications} of the {European} {Society} for the {Study} of {Science} and {Theology}, pp. 207--220. Springer International Publishing, Cham, 2020. doi: {{%
10\hspace{.1pt}\discretionary{.}{%
}{.}\hspace{.4pt}1007\discretionary{/}{%
}{/}978\discretionary{%
}{-}{-}3\discretionary{%
}{-}{-}030\discretionary{%
}{-}{-}31182\discretionary{%
}{-}{-}7\_17}}


\bibitem{hicks_learning_2020}
J.~R. Hicks and W.~P. Stewart.
\newblock Learning from wildlife-inspired awe.
\newblock {\em The Journal of Environmental Education}, 51(1), Jan. 2020.
\newblock Publisher: Routledge \_eprint: https://doi.org/10.1080/00958964.2019.1594148. doi: {{%
10\hspace{.1pt}\discretionary{.}{%
}{.}\hspace{.4pt}1080\discretionary{/}{%
}{/}00958964\hspace{.1pt}\discretionary{.}{%
}{.}\hspace{.4pt}2019\hspace{.1pt}\discretionary{.}{%
}{.}\hspace{.4pt}1594148}}


\bibitem{jung_over_2018}
S.~Jung, G.~Bruder, P.~J. Wisniewski, C.~Sandor, and C.~E. Hughes.
\newblock Over {My} {Hand}: {Using} a {Personalized} {Hand} in {VR} to {Improve} {Object} {Size} {Estimation}, {Body} {Ownership}, and {Presence}.
\newblock In {\em Proceedings of the 2018 {ACM} {Symposium} on {Spatial} {User} {Interaction}}, {SUI} '18, pp. 60--68. Association for Computing Machinery, New York, NY, USA, Oct. 2018. doi: {{%
10\hspace{.1pt}\discretionary{.}{%
}{.}\hspace{.4pt}1145\discretionary{/}{%
}{/}3267782\hspace{.1pt}\discretionary{.}{%
}{.}\hspace{.4pt}3267920}}


\bibitem{kant_critique_2009}
I.~Kant, P.~Guyer, E.~Matthews, I.~Kant, and I.~Kant.
\newblock {\em Critique of the power of judgment}.
\newblock The {Cambridge} edition of the works of {Immanuel} {Kant} / general ed.: {Paul} {Guyer} and {Allen} {W}. {Wood}. Cambridge Univ. Press, Cambridge, 11. print ed., 2009.

\bibitem{keltner_approaching_2003}
D.~Keltner and J.~Haidt.
\newblock Approaching awe, a moral, spiritual, and aesthetic emotion.
\newblock {\em Cognition and Emotion}, 17(2), Jan. 2003.
\newblock Publisher: Routledge \_eprint: https://doi.org/10.1080/02699930302297 rate: 4. doi: {{%
10\hspace{.1pt}\discretionary{.}{%
}{.}\hspace{.4pt}1080\discretionary{/}{%
}{/}02699930302297}}


\bibitem{kim_active_2023}
I.~Kim, E.~Azimi, P.~Kazanzides, and C.-M. Huang.
\newblock Active {Engagement} with {Virtual} {Reality} {Reduces} {Stress} and {Increases} {Positive} {Emotions}.
\newblock In {\em 2023 {IEEE} {International} {Symposium} on {Mixed} and {Augmented} {Reality} ({ISMAR})}, pp. 523--532, Oct. 2023.
\newblock ISSN: 2473-0726. doi: {{%
10\hspace{.1pt}\discretionary{.}{%
}{.}\hspace{.4pt}1109\discretionary{/}{%
}{/}ISMAR59233\hspace{.1pt}\discretionary{.}{%
}{.}\hspace{.4pt}2023\hspace{.1pt}\discretionary{.}{%
}{.}\hspace{.4pt}00067}}


\bibitem{kim_designing_2022}
J.~G. Kim, R.~E. Kraut, and K.~Karahalios.
\newblock Designing a {Medical} {Crowdfunding} {Website} from {Sense} of {Community} {Theory}.
\newblock {\em Proceedings of the ACM on Human-Computer Interaction}, 6(CSCW2), Nov. 2022. doi: {{%
10\hspace{.1pt}\discretionary{.}{%
}{.}\hspace{.4pt}1145\discretionary{/}{%
}{/}3555181}}


\bibitem{kitson_can_2018}
A.~Kitson and B.~E. Riecke.
\newblock Can {Lucid} {Dreaming} {Research} {Guide} {Self}-{Transcendent} {Experience} {Design} in {Virtual} {Reality}?
\newblock In {\em 2018 {IEEE} {Workshop} on {Augmented} and {Virtual} {Realities} for {Good} ({VAR4Good})}. IEEE, Reutlingen, Mar. 2018. doi: {{%
10\hspace{.1pt}\discretionary{.}{%
}{.}\hspace{.4pt}1109\discretionary{/}{%
}{/}VAR4GOOD\hspace{.1pt}\discretionary{.}{%
}{.}\hspace{.4pt}2018\hspace{.1pt}\discretionary{.}{%
}{.}\hspace{.4pt}8576889}}


\bibitem{kitson_are_2018}
A.~Kitson, T.~Schiphorst, and B.~E. Riecke.
\newblock Are {You} {Dreaming}? {A} {Phenomenological} {Study} on {Understanding} {Lucid} {Dreams} as a {Tool} for {Introspection} in {Virtual} {Reality}.
\newblock In {\em Proceedings of the 2018 {CHI} {Conference} on {Human} {Factors} in {Computing} {Systems}}, {CHI} '18, pp. 1--12. Association for Computing Machinery, New York, NY, USA, Apr. 2018. doi: {{%
10\hspace{.1pt}\discretionary{.}{%
}{.}\hspace{.4pt}1145\discretionary{/}{%
}{/}3173574\hspace{.1pt}\discretionary{.}{%
}{.}\hspace{.4pt}3173917}}


\bibitem{kitson_designing_2020}
A.~Kitson, E.~R. Stepanova, I.~A. Aguilar, N.~Wainwright, and B.~E. Riecke.
\newblock Designing {Mind}(set) and {Setting} for {Profound} {Emotional} {Experiences} in {Virtual} {Reality}.
\newblock In {\em Proceedings of the 2020 {ACM} {Designing} {Interactive} {Systems} {Conference}}, {DIS} '20, pp. 655--668. Association for Computing Machinery, New York, NY, USA, July 2020. doi: {{%
10\hspace{.1pt}\discretionary{.}{%
}{.}\hspace{.4pt}1145\discretionary{/}{%
}{/}3357236\hspace{.1pt}\discretionary{.}{%
}{.}\hspace{.4pt}3395560}}


\bibitem{kokkinara_measuring_2014}
E.~Kokkinara and M.~Slater.
\newblock Measuring the {Effects} through {Time} of the {Influence} of {Visuomotor} and {Visuotactile} {Synchronous} {Stimulation} on a {Virtual} {Body} {Ownership} {Illusion}.
\newblock {\em Perception}, 43(1):43--58, Jan. 2014.
\newblock Publisher: SAGE Publications Ltd STM. doi: {{%
10\hspace{.1pt}\discretionary{.}{%
}{.}\hspace{.4pt}1068\discretionary{/}{%
}{/}p7545}}


\bibitem{liu_virtual_2022}
P.~Liu, E.~R. Stepanova, A.~Kitson, T.~Schiphorst, and B.~E. Riecke.
\newblock Virtual {Transcendent} {Dream}: {Empowering} {People} through {Embodied} {Flying} in {Virtual} {Reality}.
\newblock In {\em {CHI} {Conference} on {Human} {Factors} in {Computing} {Systems}}, pp. 1--18. ACM, New Orleans LA USA, Apr. 2022. doi: {{%
10\hspace{.1pt}\discretionary{.}{%
}{.}\hspace{.4pt}1145\discretionary{/}{%
}{/}3491102\hspace{.1pt}\discretionary{.}{%
}{.}\hspace{.4pt}3517677}}


\bibitem{longinus_sublime_1995}
Longinus, Aristotle, and Demetrius.
\newblock On the {Sublime}.
\newblock In {\em Aristotle: {Poetics}; {Longinus}: {On} the {Sublime}; {Demetrius}: {On} {Style}}, number L199 in Loeb classical library. Harvard University Press, Cambridge, Mass, 1995.

\bibitem{miller_awedyssey_2023}
N.~Miller, E.~R. Stepanova, J.~Desnoyers-Stewart, A.~Adhikari, A.~Kitson, P.~Pennefather, D.~Quesnel, K.~Brauns, A.~Friedl-Werner, A.~Stahn, and B.~E. Riecke.
\newblock Awedyssey: {Design} {Tensions} in {Eliciting} {Self}-transcendent {Emotions} in {Virtual} {Reality} to {Support} {Mental} {Well}-being and {Connection}.
\newblock In {\em Proceedings of the 2023 {ACM} {Designing} {Interactive} {Systems} {Conference}}, {DIS} '23. Association for Computing Machinery, New York, NY, USA, July 2023. doi: {{%
10\hspace{.1pt}\discretionary{.}{%
}{.}\hspace{.4pt}1145\discretionary{/}{%
}{/}3563657\hspace{.1pt}\discretionary{.}{%
}{.}\hspace{.4pt}3595998}}


\bibitem{monroy_awe_2022}
M.~Monroy and D.~Keltner.
\newblock Awe as a {Pathway} to {Mental} and {Physical} {Health}.
\newblock {\em Perspectives on Psychological Science}, Aug. 2022.
\newblock Publisher: SAGE Publications Inc. doi: {{%
10\hspace{.1pt}\discretionary{.}{%
}{.}\hspace{.4pt}1177\discretionary{/}{%
}{/}17456916221094856}}


\bibitem{monroy_influences_2023}
M.~Monroy, Ã.~Uğurlu, F.~Zerwas, R.~Corona, D.~Keltner, J.~Eagle, and M.~Amster.
\newblock The influences of daily experiences of awe on stress, somatic health, and well-being: a longitudinal study during {COVID}-19.
\newblock {\em Scientific Reports}, 13(1):9336, June 2023.
\newblock Number: 1 Publisher: Nature Publishing Group. doi: {{%
10\hspace{.1pt}\discretionary{.}{%
}{.}\hspace{.4pt}1038\discretionary{/}{%
}{/}s41598\discretionary{%
}{-}{-}023\discretionary{%
}{-}{-}35200\discretionary{%
}{-}{-}w}}


\bibitem{moreira-almeida_wpa_2016}
A.~Moreira-Almeida, A.~Sharma, B.~J. van Rensburg, P.~J. Verhagen, and C.~C. Cook.
\newblock {WPA} {Position} {Statement} on {Spirituality} and {Religion} in {Psychiatry}.
\newblock {\em World Psychiatry}, 15(1), 2016.
\newblock \_eprint: https://onlinelibrary.wiley.com/doi/pdf/10.1002/wps.20304. doi: {{%
10\hspace{.1pt}\discretionary{.}{%
}{.}\hspace{.4pt}1002\discretionary{/}{%
}{/}wps\hspace{.1pt}\discretionary{.}{%
}{.}\hspace{.4pt}20304}}


\bibitem{mossbridge_designing_2016}
J.~Mossbridge.
\newblock Designing transcendence technology.
\newblock {\em Psychology's New Design Science and the Reflective Practitioner}, 2016.
\newblock Publisher: LibraLab Press.

\bibitem{ogawa_virtual_2019}
N.~Ogawa, T.~Narumi, and M.~Hirose.
\newblock Virtual {Hand} {Realism} {Affects} {Object} {Size} {Perception} in {Body}-{Based} {Scaling}.
\newblock In {\em 2019 {IEEE} {Conference} on {Virtual} {Reality} and {3D} {User} {Interfaces} ({VR})}, pp. 519--528, Mar. 2019.
\newblock ISSN: 2642-5254. doi: {{%
10\hspace{.1pt}\discretionary{.}{%
}{.}\hspace{.4pt}1109\discretionary{/}{%
}{/}VR\hspace{.1pt}\discretionary{.}{%
}{.}\hspace{.4pt}2019\hspace{.1pt}\discretionary{.}{%
}{.}\hspace{.4pt}8798040}}


\bibitem{peck_avatar_2021}
T.~C. Peck and M.~Gonzalez-Franco.
\newblock Avatar {Embodiment}. {A} {Standardized} {Questionnaire}.
\newblock {\em Frontiers in Virtual Reality}, 1, Feb. 2021. doi: {{%
10\hspace{.1pt}\discretionary{.}{%
}{.}\hspace{.4pt}3389\discretionary{/}{%
}{/}frvir\hspace{.1pt}\discretionary{.}{%
}{.}\hspace{.4pt}2020\hspace{.1pt}\discretionary{.}{%
}{.}\hspace{.4pt}575943}}


\bibitem{pizarro_self-transcendent_2021}
J.~J. Pizarro, N.~Basabe, {Itziar Fernández}, P.~Carrera, P.~Apodaca, C.~I. Man~Ging, O.~Cusi, and D.~Páez.
\newblock Self-{Transcendent} {Emotions} and {Their} {Social} {Effects}: {Awe}, {Elevation} and {Kama} {Muta} {Promote} a {Human} {Identification} and {Motivations} to {Help} {Others}.
\newblock {\em Frontiers in Psychology}, 12, 2021. doi: {{%
10\hspace{.1pt}\discretionary{.}{%
}{.}\hspace{.4pt}3389\discretionary{/}{%
}{/}fpsyg\hspace{.1pt}\discretionary{.}{%
}{.}\hspace{.4pt}2021\hspace{.1pt}\discretionary{.}{%
}{.}\hspace{.4pt}709859}}


\bibitem{quesnel_are_2018}
D.~Quesnel and B.~E. Riecke.
\newblock Are {You} {Awed} {Yet}? {How} {Virtual} {Reality} {Gives} {Us} {Awe} and {Goose} {Bumps}.
\newblock {\em Frontiers in Psychology}, 9, 2018.

\bibitem{quesnel_creating_2018}
D.~Quesnel, E.~R. Stepanova, I.~A. Aguilar, P.~Pennefather, and B.~E. Riecke.
\newblock Creating {AWE}: {Artistic} and {Scientific} {Practices} in {Research}-{Based} {Design} for {Exploring} a {Profound} {Immersive} {Installation}.
\newblock In {\em 2018 {IEEE} {Games}, {Entertainment}, {Media} {Conference} ({GEM})}, pp. 1--207, Aug. 2018. doi: {{%
10\hspace{.1pt}\discretionary{.}{%
}{.}\hspace{.4pt}1109\discretionary{/}{%
}{/}GEM\hspace{.1pt}\discretionary{.}{%
}{.}\hspace{.4pt}2018\hspace{.1pt}\discretionary{.}{%
}{.}\hspace{.4pt}8516463}}


\bibitem{r_core_team_r_2013}
R.~R~Core~Team and {others}.
\newblock R: {A} language and environment for statistical computing.
\newblock 2013.
\newblock Publisher: Vienna, Austria.

\bibitem{ries_effect_2008}
B.~Ries, V.~Interrante, M.~Kaeding, and L.~Anderson.
\newblock The effect of self-embodiment on distance perception in immersive virtual environments.
\newblock In {\em Proceedings of the 2008 {ACM} symposium on {Virtual} reality software and technology}, {VRST} '08, pp. 167--170. Association for Computing Machinery, New York, NY, USA, Oct. 2008. doi: {{%
10\hspace{.1pt}\discretionary{.}{%
}{.}\hspace{.4pt}1145\discretionary{/}{%
}{/}1450579\hspace{.1pt}\discretionary{.}{%
}{.}\hspace{.4pt}1450614}}


\bibitem{riva_neuroscience_2019}
G.~Riva, B.~K. Wiederhold, and F.~Mantovani.
\newblock Neuroscience of {Virtual} {Reality}: {From} {Virtual} {Exposure} to {Embodied} {Medicine}.
\newblock {\em Cyberpsychology, Behavior, and Social Networking}, 22(1):82--96, Jan. 2019.
\newblock Publisher: Mary Ann Liebert, Inc., publishers. doi: {{%
10\hspace{.1pt}\discretionary{.}{%
}{.}\hspace{.4pt}1089\discretionary{/}{%
}{/}cyber\hspace{.1pt}\discretionary{.}{%
}{.}\hspace{.4pt}2017\hspace{.1pt}\discretionary{.}{%
}{.}\hspace{.4pt}29099\hspace{.1pt}\discretionary{.}{%
}{.}\hspace{.4pt}gri}}


\bibitem{schopenhauer_world_1969}
A.~Schopenhauer and A.~Schopenhauer.
\newblock {\em The world as will and representation (volume 1)}.
\newblock Dover Pub, New York, 1969.

\bibitem{silvia_openness_2015}
P.~J. Silvia, K.~Fayn, E.~C. Nusbaum, and R.~E. Beaty.
\newblock Openness to experience and awe in response to nature and music: {Personality} and profound aesthetic experiences.
\newblock {\em Psychology of Aesthetics, Creativity, and the Arts}, 9, 2015.
\newblock Place: US Publisher: Educational Publishing Foundation. doi: {{%
10\hspace{.1pt}\discretionary{.}{%
}{.}\hspace{.4pt}1037\discretionary{/}{%
}{/}aca0000028}}


\bibitem{somarathna_virtual_2022}
R.~Somarathna, T.~Bednarz, and G.~Mohammadi.
\newblock Virtual {Reality} for {Emotion} {Elicitation} – {A} {Review}.
\newblock {\em IEEE Transactions on Affective Computing}, pp. 1--21, 2022.
\newblock Conference Name: IEEE Transactions on Affective Computing. doi: {{%
10\hspace{.1pt}\discretionary{.}{%
}{.}\hspace{.4pt}1109\discretionary{/}{%
}{/}TAFFC\hspace{.1pt}\discretionary{.}{%
}{.}\hspace{.4pt}2022\hspace{.1pt}\discretionary{.}{%
}{.}\hspace{.4pt}3181053}}


\bibitem{stepanova_spacevirtual_2019}
E.~R. Stepanova, D.~Quesnel, and B.~E. Riecke.
\newblock Space—{A} {Virtual} {Frontier}: {How} to {Design} and {Evaluate} a {Virtual} {Reality} {Experience} of the {Overview} {Effect}.
\newblock {\em Frontiers in Digital Humanities}, 6, 2019. doi: {{%
10\hspace{.1pt}\discretionary{.}{%
}{.}\hspace{.4pt}3389\discretionary{/}{%
}{/}fdigh\hspace{.1pt}\discretionary{.}{%
}{.}\hspace{.4pt}2019\hspace{.1pt}\discretionary{.}{%
}{.}\hspace{.4pt}00007}}


\bibitem{stepanova_understanding_2019}
E.~R. Stepanova, D.~Quesnel, and B.~E. Riecke.
\newblock Understanding {AWE}: {Can} a {Virtual} {Journey}, {Inspired} by the {Overview} {Effect}, {Lead} to an {Increased} {Sense} of {Interconnectedness}?
\newblock {\em Frontiers in Digital Humanities}, 6, 2019. doi: {{%
10\hspace{.1pt}\discretionary{.}{%
}{.}\hspace{.4pt}3389\discretionary{/}{%
}{/}fdigh\hspace{.1pt}\discretionary{.}{%
}{.}\hspace{.4pt}2019\hspace{.1pt}\discretionary{.}{%
}{.}\hspace{.4pt}00009}}


\bibitem{valdesolo_awe_2014}
P.~Valdesolo and J.~Graham.
\newblock Awe, {Uncertainty}, and {Agency} {Detection}.
\newblock {\em Psychological Science}, 25(1), Jan. 2014.
\newblock Publisher: SAGE Publications Inc. doi: {{%
10\hspace{.1pt}\discretionary{.}{%
}{.}\hspace{.4pt}1177\discretionary{/}{%
}{/}0956797613501884}}


\bibitem{yaden_development_2019}
D.~B. Yaden, S.~B. Kaufman, E.~Hyde, A.~Chirico, A.~Gaggioli, J.~W. Zhang, and D.~Keltner.
\newblock The development of the {Awe} {Experience} {Scale} ({AWE}-{S}): {A} multifactorial measure for a complex emotion.
\newblock {\em The Journal of Positive Psychology}, 14(4), July 2019.
\newblock Publisher: Routledge \_eprint: https://doi.org/10.1080/17439760.2018.1484940. doi: {{%
10\hspace{.1pt}\discretionary{.}{%
}{.}\hspace{.4pt}1080\discretionary{/}{%
}{/}17439760\hspace{.1pt}\discretionary{.}{%
}{.}\hspace{.4pt}2018\hspace{.1pt}\discretionary{.}{%
}{.}\hspace{.4pt}1484940}}


\end{thebibliography}

\newcommand{\originalthesection}{\thesection}
\renewcommand{\thesection}{A}
\section{Semi-Structured Interview Questions}

How do you feel right now? What part of the scene left an impression on you or evoked certain feelings? \\
Can you describe your overall feeling about the experience? \\
In the scene without the moon and waves, what did you feel? In the scene with the moon and waves, what did you feel? What are the differences between them? \\
How did you feel when it turned to night? After the moon set and the sun rose, what were you thinking? \\
Have you experienced this feeling in your daily life? Can you describe such a scene? \\
During the rise, movement, and disappearance of the giant wave and moon, what were you thinking? How did you feel? \\
Can you comment on the sounds of the seaside and the giant wave you heard? How did they make you feel? \\
Regarding the scene's sound effects, visuals, interactions, etc., did you have any expectations during the experience that weren't met? \\
During the experience, how did you perceive your real body? \\
Did you notice your virtual body during the process? What effect did it have? \\
When you felt the mist of water on your hand, how did your body feel? What were you thinking at that moment? \\
Did you notice the shadow on the beach when the moon set and the sun rose? What did you do then? How did you feel?

\renewcommand{\thesection}{B}
\section{Part of The Referenced Theory of Concepts}
\renewcommand{\thesection}{\originalthesection}

\begin{table}[h]
  \caption{\centering High-level concepts, middle-level concepts, and part of the referenced theory.}
  \label{tab2}
  \scriptsize%
  \centering%
  \begin{tabu} to \columnwidth {%
    X[1,l]  
    X[1.5,l]
    X[3,l]
  }
    \toprule
    High-level concepts & Middle-level concepts & Part of the referenced theory \\
    \midrule
    \multirow{4}{*}{Terror} & Power & A superior power that is superior to the individual to sublime~\cite{schopenhauer_world_1969}.\\
    & Excessive loudness & Excessive loudness alone is sufficient to overpower the soul~\cite{burke_philosophical_2019}. \\
    & Painful sounds & Sounds that mimic animals in pain or danger can convey great ideas~\cite{burke_philosophical_2019}. \\
    & Magnitude & Magnitude refers to something that is absolutely great, that is, sublime~\cite{kant_critique_2009}. \\
    \multirow{2}{*}{Distance or modification} & Safety & Actual danger disrupts peace, leading to the diminishment of the sublime~\cite{schopenhauer_world_1969}.\\
    & Understandability & An object should not be monstrous or colossal because it would overwhelm our ability to understand it~\cite{kant_critique_2009}. \\
    \multirow{6}{*}{Obscurity} & Unfamiliar situations & Knowing the full extent of and getting used to danger diminishes the sublime~\cite{burke_philosophical_2019}.\\
    & Extreme darkness or light & To make an object striking and evoke the sublime, it should contrast sharply with its surroundings~\cite{burke_philosophical_2019}. \\
    & Sad and fuscous colors & The sublime materials should be in dark and somber colors like black, brown, deep purple, and similar tones~\cite{burke_philosophical_2019}. \\
    & Suddenness & When something sudden and unexpected occurs, we perceive danger~\cite{burke_philosophical_2019}.\\
    & Intermitting & A flickering light is more terrifying than total darkness~\cite{burke_philosophical_2019}.\\
    & Quick transition & A rapid shift from light to darkness or vice versa has a great impact~\cite{burke_philosophical_2019}. \\
    \bottomrule
  \end{tabu}
\end{table}

\end{document}